\documentclass[usenatbib]{mn2e}
\input epsf

\newcommand{\rd}{{\rm d}}
\newcommand{\beq}{\begin{equation}}
\newcommand{\eeq}{\end{equation}}
\newcommand{\beqary}{\begin{eqnarray}}
\newcommand{\eeqary}{\end{eqnarray}}

\title[Millisecond dips in the RXTE/PCA lightcurve of Sco X-1]{Millisecond 
dips in the
2007-2009 RXTE/PCA lightcurve of Sco X-1 and one possible occultation 
event}
\author[Chang, Liu, and Chen]{Hsiang-Kuang Chang$^{1,2}$\thanks{E-mail:
hkchang@phys.nthu.edu.tw}, Chih-Yuan Liu$^2$, and Kuan-Ting Chen$^2$ 
\\
$^{1}$Institute of Astronomy, National Tsing Hua University, 
Hsinchu 30013, Taiwan\\ 
$^{2}$Department of Physics, National Tsing Hua University, 
Hsinchu 30013, Taiwan} 

\begin{document}

\date{July 2010; September 2010}

\pagerange{\pageref{firstpage}--\pageref{lastpage}} \pubyear{2010}

\maketitle

\label{firstpage}

\begin{abstract}
Serendipitous stellar occultation search is so far the only way 
to detect the existence of very small, very dim, remote objects 
in the solar system. 
To date, however, there are only very few 
reported detections for trans-Neptunian objects (TNOs) in optical bands. 
In the X-ray band, with the RXTE/PCA data of Sco X-1 taken from
June 2007 to October 2009, we found one possible X-ray occultation event. 
We discuss the veracity and properties of this event, and suggest
upper limits to the size distribution of TNOs at hectometer size
and of Main-Belt Asteroids (MBAs)
at decameter size.  
\end{abstract}


\section{Introduction}
Population properties of small solar-system bodies, 
including main-belt asteroids (MBAs) and trans-Neptunian objects (TNOs), 
among others, carry information of the early solar system 
when planets were being formed and of the physics 
in the dynamical and collisional history of our solar system
(see, for example, \citet{kenyon08,bottke05,obrien05,cheng04}).
Our knowledge of them, however, is far from being complete, 
particularly for those very small ones. 
Direct observation is so far achieved only for TNOs 
larger than deca-kilometers and for MBAs larger than about 400 meters. 
To pin down their size distribution at the small size end, 
searching for serendipitous stellar occultation events was proposed 
to be a possible way 
\citep{bailey76,brown97,roques00,cooray03}.
Such search has been being conducted mainly 
in optical bands 
\citep{bianco10,bickerton08,roques03}
with few reported detections
\citep{roques06,schlichting09}. 
In the X-ray band, 
in which occultation by even smaller objects may be more easily 
detected because of a smaller Fresnel scale, 
putative occultation events, allegedly caused by 100-meter size TNOs, 
were found in the 1996-2002 X-ray data of Sco X-1 taken 
by the instrument Proportional Counter Array (PCA)
on board Rossi X-ray Timing Explorer (RXTE)
\citep{chang06,chang07}. 
Those events, however, were later found 
to be most likely related to some instrumental dead-time effect 
of unknown nature
\citep{chang07,jones08,liu08,blocker09}. 
To get rid of instrumental effect contamination,
new RXTE/PCA observations of Sco X-1 with a newly designed data mode 
to record detailed information of each PCA-detected high energy event 
have been being conducted since June 2007.  
It is hoped that those high energy events, 
which are called `very large events' (VLEs) in RXTE/PCA terminology, 
can provide clear signature of possible instrumental effects. 

In this paper we report the result we obtained 
from analyzing the new data taken from June 2007 to October 2009, 
which in total is about 240 ks.
We found that VLEs, classified into different types according to
the number of triggered anodes recorded in the data,
can be used 
as an indicator of instrumental effects. 
In the 240-ks data, only 1 dip event is found to be unrelated to
VLEs. We discuss the property of this dip event
with occultation 
diffraction pattern fitting to its light curve. 
We then estimate upper limits 
to the size distribution of TNOs at hectometer size
and of MBAs at decameter size, which are unattainable by any other means
known to date. 

\section{Dip events and the RXTE/PCA VLEs}

The new RXTE observation of Sco X-1 started from 
June 13, 2007, and is still going on (RXTE ObsId 93067).
Technical details of RXTE instrumentation can be found in the RXTE web site and also in 
\citet{jahoda06}.
To search for millisecond dip events, 
we adopted the same procedure employed in previous studies 
\citep{chang06,chang07,liu08}.
We examined the light curves of Sco X-1 
binned in different bin sizes from 1 ms to 10 ms.
The deviation in photon counts of each bin in the light curve was 
determined by comparing its count number with the average counts 
and the count variance in an 8-s running window.
In the 240-ks data that we use for this analysis, 
39 `significant' and 253 `less significant' dips are found. 
To compare with our earlier works, we define `significant' 
as those dips found in our searching algorithm with negative 
deviation larger than 6.5 $\sigma$ and `less significant' 
as that between 5 $\sigma$ and 6.5 $\sigma$. 
The choice of (minus) 6.5 $\sigma$ in previous works was to set 
a random probability at 0.001 for the analysis of the original 320-ks data
employed in \citet{chang06},
and (minus) 5 $\sigma$ is about the level 
below which the deviation distribution
obviously shows excess compared to a normal distribution.
We keep the same choice of the deviation level
as in earlier works to have a better comparison.
Readers are referred to 
\citet{chang06,chang07,liu08}.
for details of our searching algorithm.
To investigate the connection between these dip events
and possible instrumental effects, we examined detailed information of 
the recorded VLEs as described below.

RXTE/PCA VLEs are those events which deposit more than 100 keV 
in any one of the 6 active xenon anodes and the propane anode 
of one of the five identical proportional counter units (PCUs) of PCA. 
A VLE saturates
the pre-amplifier and causes ringing in the signal chain when
the amplified signal is restored to the baseline.
Longer system dead time is set for VLEs, 
usually  at 50 $\mu$s for observations of Sco X-1. 
The actual
duration in which the signal chains are affected depends on the
pulse height of the saturating event.
In default setting of RXTE/PCA observations, 
VLE count rates are recorded in a standard data mode 
with 125-ms time resolution. 
In the new observation of Sco X-1 (RXTE ObsID: 93067), 
to explore possible instrumental effects which may cause 
the millisecond dips in the PCA light curve of Sco X-1, 
a particular data mode was designed to 
retain detailed information of each individual VLE. 
In this data mode, the event epoch is recorded with 125-$\mu$s 
resolution and the identification of anodes triggered 
by the VLE is also recorded. 
With this new data mode, we found that some VLEs are very unusual 
in the way that no anode is recorded as having been triggered. 
We therefore classify VLEs into 4 types according to the number 
of triggered anodes as listed in Table~\ref{vletype}, 
in which the averaged count rates (in units of counts per second per PCU) 
of these different VLEs are also listed. 
The count rate for all VLEs is 90.5$\pm$19.2 counts per second per PCU. 
It is clear that Type-A VLEs are rare.
\begin{table}
\begin{center}
\begin{tabular}{clc}
\hline
Type & number of triggered anodes & average count rate \\
\hline
A & no  & $1.73\pm 0.29$\\ 
B & all  & $34.5\pm 11.2$\\ 
C & more than one but not all & $44.8\pm 9.80$\\ 
D & only one & $9.44\pm 2.27$\\ 
\hline
\end{tabular}
\end{center}
\caption{VLE types and their average count rates (in units of counts 
per second per PCU). The number of PCUs which are on during the observation 
varies from time to time. The total length of the data employed 
in this analysis is 240 ksec, and is 970 ksec-PCU when the number of PCU on
is taken into account. } 
\label{vletype}
\end{table}  

To study the association of the dip events that we found with VLEs, 
we classify the dip events into 5 groups according to the type of VLEs 
that were detected within the time interval defined by the dip duration 
and one duration before the dip epoch. 
The number of dips in different groups is shown in Table~\ref{dipgrp}. 
It is obvious that dip events are strongly related to Type-A VLEs. 
Among the 39 significant dips 34 are associated with Type-A VLEs, 
which are rare among all VLEs. 
It is not clear yet why a VLE is recorded in the data with no anode triggered.
VLEs are usually caused by high energy particles. 
It may be possible that the incident particle is very energetic 
and an ionization avalanche or a secondary-particle shower 
is created so that the instrument suffers from a long sequence of dead time
\citep{jones08}. 
For some unknown reasons all the VLE triggering flags are turned off. 
Although not yet fully understood, we conclude that Group-A dip events 
are instrumental.
\begin{table}
\begin{center}
\begin{tabular}{llcc}
\hline
Group &  associated VLE type  & significant  & less significant  \\
 & & dips & dips \\
\hline
A & Type A & 34 & 125 \\
B & Type B, no Type A & 2 & 94 \\
C & Type C, no Type A, B & 2 & 23 \\
D & Type D only & 0 & 3 \\
E & no VLEs & 1 & 8 \\ 
\hline
\end{tabular}
\end{center}
\caption{Number of dips in different groups.}
\label{dipgrp}
\end{table}  

\begin{figure}
\epsfxsize=8.4cm
\epsffile{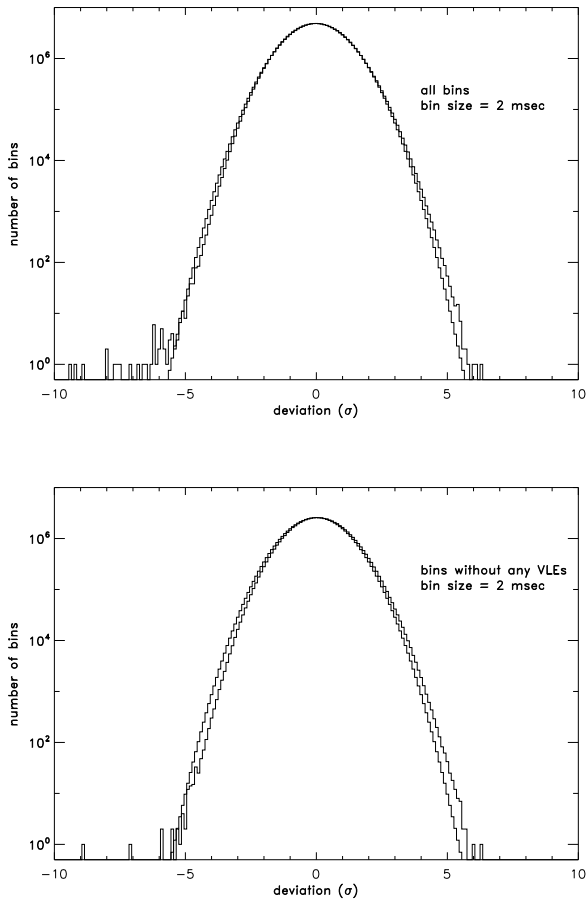}
\caption{Deviation distribution of the Sco X-1 RXTE/PCA light curve 
based on the 240-ks data taken from June 2007 to October 2009 that we employed
in this work. Thick histograms are from the data with 2-ms time bins, and thin
histograms are the corresponding Gaussion distribution 
with the same total number of bins. 
The upper panel is for all the time bins, and the lower panel is
for only those bins within which there are no VLEs.
In the lower panel, most of those bins with large negative deviation,
except for the one at about $-7.1 \sigma$, are related to dip events
of Group A-D, although within the 2-ms bin itself there is no VLE. 
}
\label{devdis}
\end{figure}

To investigate whether dip events in other groups are also 
instrumental, we examine the consistency between the number
of less significant dips and that due to random fluctuation.
The deviation distribution of all the time bins in the light curve 
should be of a Poisson nature modified by the dead-time 
and coincidence-event effects 
if all the fluctuations are random 
and there is no occultation event. 
One example is shown in Figure~\ref{devdis} for 
a search with 2-ms time bins.
Similar figures for earlier data can be found in 
\citet{chang06,chang07,liu08}.
From Figure~\ref{devdis} we can see that, although 
Gaussian distribution is a good approximation, the deviation
distribution of the data is about a factor of $2\sim 3$ lower than
the Gaussian one at $-5\sigma$. 
In our current discussion we mainly consider the search with 2-ms bins since
most of the less significant dips have a duration of 2 ms. 
If a Gaussian distribution is assumed, one expects to have 34.4 bins
with photon counts less than 5 $\sigma$ below the average 
for 2-ms bins in the 240-ks data.
Since some VLEs happen at the same time (within the $125\mu$s time resolution)
or within the same 2-ms bin, to estimate the distribution of these 34.4 bins
in different dip groups, we first count the number of bins without any VLEs.
The total amount of those bins is about 64.5 ks in the 120-ks data.
Therefore, 18.5 of the 34.4 bins should be associated with Group E.
For the other dip groups, we distribute the remaining 15.9 bins according
to their related VLE count rates. 
The result is that, assuming a Gaussian distribution,
there should be 0.3, 6.1, 7.9, 1.7, and 18.5 bins associated with
Group A to E respectively.
We note that this is in fact not strictly correct, because
as mentioned above some VLEs happen in the same 2-ms time bin and
we assign dip events into different groups 
with highest priority for Type-A VLEs, that is, there can be Type-B VLEs
associated with Group-A dips.
Nontheless, the information here is 
enough for us to draw conclusions.

\begin{figure}
\epsfxsize=8.4cm
\epsffile{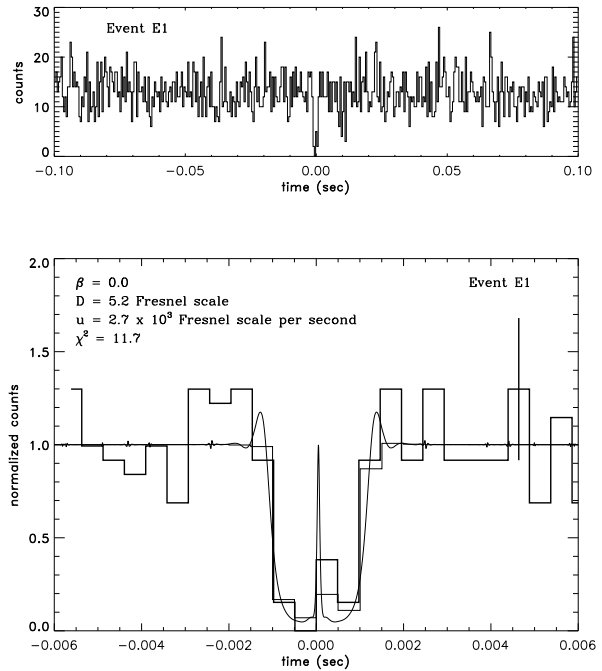}
\caption{Light curves of Event E1. 
The upper panel shows the light curve 100-ms before and after Event E1
in 0.5-ms bins.
The lower panel is a zoom-in view of Event E1.
The thick histogram is the observed light curves in 0.5-ms bins, 
the thin curve is the computed diffraction pattern 
with best-fit parameters, and the thin histogram is the binned light curve 
of the computed one. 
The level of one-$\sigma$ uncertainty in the observed light curve 
is indicated with a vertical bar. 
The average number of counts in a 0.5-ms
bin is 13.1.  
}
\label{lc}
\end{figure}

Comparing the numbers of less significant dips in different groups
with those derived from a Gaussian distribution, and further taking
in to account the reduction by a factor of $2\sim 3$,
one can see that the numbers of less significant dips of Group A, B,
C and D are not consistent with random fluctuation.
There is a tendancy that the excess of dip numbers 
decreases from Group A to Group D. 
Although details are not yet fully understood, it seems clear that
Type-A VLEs are the most energetic and Type-D VLEs are the weakest. 
Because of the association with VLEs, all the Group A-D dips that are not
due to random fluctuation are most likely instrumental.
On the other hand, the number of less significant dips of Group E is roughly
consistent with random fluctuation. 
We have no indication of instrumental effects for Group-E dips, 
which are not associated with any VLEs.
The significant dip in Group E, which is found in the search with
2-ms bins, 
has a deviation of $-7.1 \sigma$,
corresponding to a random probability of less than $7.5 \times 10^{-5}$ in
the 240-ks data. 
We denote this dip event as Event E1, whose
light curve is plotted in Figure~\ref{lc},
together with the best-fit diffraction pattern described in the next section.

\section{Distance and size of the occulting bodies}

Event E1, a non-instrumental millisecond dip event, 
is most likely an occultation event
caused by objects in the solar system
\citep{chang06,chang07}. The observed light curve, indeed, can be well
described by shadows of occultation when diffraction 
is properly taken into account. In the following we explain how we
conduct the diffraction pattern fitting and how we estimate the
distance and size of the occulting body with fitting results and 
assumed orbital parameters.
  
\subsection{Fitting the occultation light curve with diffraction patterns}

In our study we assume the occulting body is a spherical one and
the background star, Sco X-1, is a distant point source.
We also take into account the actual RXTE/PCA observed Sco X-1 spectrum,
which peaks at 4 keV.  
We note that the shadow diffraction patterns are all  the same for a
given ratio of the occulting body diameter to the Fresnel scale when
the length is expressed in units of the Fresnel scale.
Hereafter we denote that ratio with $\alpha$.
To describe an occultation light curve, 4 more parameters are needed.
One is the impact parameter $\beta$, 
which is defined as 
the closest distance of the crossing path to the shadow center
in units of the occulting body radius.
Another one is the transverse relative speed $u_{\rm t}$ of the observer 
with respect to the shadow. 
The third is the epoch of the central crossing, $t_0$, and
the fourth is the nomalization $N$.
For each occultation light curve, to find a best fit in the
5-parameter space ($\alpha$, $\beta$, $u_{\rm t}$, $t_0$, $N$)
is not practical.
We therefore search for the `best' fit in the following way.
We fix $\beta$ at certain different values,
e.g., from 0 to 1 with an increment step of 0.1.
For a fixed $\beta$, we compute theoretical occultation light curves 
over a range of $\alpha$. For each computed light curve of a given $\alpha$,
$u_{\rm t}$ over a certain range  is applied to convert the length 
of the computed pattern into time and the computed curve is folded into
a binned curve to compare with data.

In this study, we use data curves with 0.5-ms bins.
The normalization $N$ 
of the computed curve is set to be the average counts per bin
of the data curve in the fitting window, which
we choose to be 12.5 ms, that is, 25 bins.
With given $\beta$, $\alpha$, $u_{\rm t}$, and $N$,
we adjust $t_0$ with a step size of 0.05 ms, which is 0.1 time bin,
to look for the smallest $\chi^2$ between the binned curve and data curve.
We then assign this $\chi^2$ to be the fitting result of this
set of parameters $\beta$, $\alpha$, and $u_{\rm t}$. 
The `best' fit is determined by the minimum $\chi^2$ in the 
parameter space of  $\beta$, $\alpha$, and $u_{\rm t}$ as 
described above. 

\begin{figure}
\epsfxsize=8.4cm
\epsffile{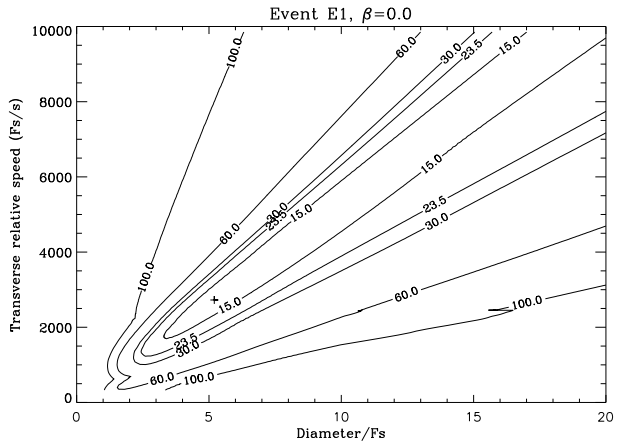}
\caption{$\chi^2$ contours of the diffraction pattern 
fitting in the $\alpha$-$u_{\rm t}$ parameter
space for Event E1. The cross symbol marks the location
where the smallest $\chi^2$ value is found.
}
\label{cont}
\end{figure}

The best fit result for Event E1 is that  
$\chi^2=11.7$ (with 20 degrees of freedom),  
$\beta=0.0$, $\alpha=5.2_{-2.6}$ and 
$u_{\rm t}=2700_{-1450}$ Fresnel scale per second.
The uncertainty quoted above is at the 3-$\sigma$ level
in the $\alpha$-$u_{\rm t}$ space for a fixed $\beta$. 
Here we report only the lower uncertainty level, 
since a combination of large $u_{\rm t}$ and large $\alpha$ gives
acceptable fits; see Figure~\ref{cont}.
The reason is that
the light curve of Event E1 does not show clear side lobes.
Diffraction patterns of large $u_{\rm t}$ and $\alpha$ therefore
cannot be discriminated. 
Those patterns correspond to occultation caused by very nearby objects
(see the next subsection).
The uncertainty in $\alpha$ quoted here is the value corresponding
to the uncertainty in $u_{\rm t}$, rather than the $\alpha$
extreme values on the 3-$\sigma$ level contour.
It is because that our interest in $\alpha$ is
for estimating the diameter of the occulting body,
which depends on both $u_{\rm t}$ and $\alpha$ in a
non-trivial way, and the distance estimate
depends on $u_{\rm t}$ only.
The $\beta$ dependence of the diffraction pattern fitting is weak.
The best-fit $\chi^2$ value (11.7 for 20 d.o.f.) is very low. 
It only shows that the light curve can be described by a diffraction
pattern, but at the same time provides less power in discriminating
fitting parameters.
This low $\chi^2$ value are due the following two factors.
One is the low count number in each bin, which gives a relatively large
error bar, and the other is the 0.5-ms bin size, which is large so that
details of a diffraction pattern are smeared out.
These two factors make it easy to obtain a low $\chi^2$ value fitting. 

\subsection{Distance and size determination assuming a circular orbit}

\begin{figure}
\epsfxsize=8.4cm
\epsffile{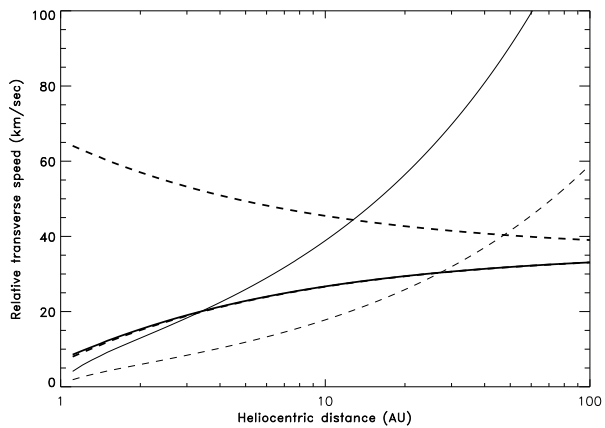}
\caption{Estimate of the occulting body distance for Event E1. 
The thick curves are the relative speed transverse 
to the Sco X-1 direction between RXTE and the occulting body, 
whose orbit is assumed to be a circular one. 
The thick solid curve is for the case of a minimum orbital inclination. 
The thick dashed curves give the largest 
and smallest relative transverse speed 
when all the possible orbital inclination is considered. 
The lower thick dashed curve almost coincident 
with the thick solid one. 
In this computation, RXTE velocity in the solar system 
at the event epoch and the Sco X-1 direction, 
which is 5.5 degrees north of the ecliptic, are both taken into account.  
The thin curves are the relative transverse speed determined 
from diffraction pattern fitting. 
The thin solid one is the best fit 
and the thin dashed one give the uncertainty at the 3-$\sigma$ level.  
For Event E1, diffraction pattern fitting does not provide 
good constraint to the higher end of the relative speed, 
because of lacking of side lobes. 
}
\label{dist}
\end{figure}

Based on the best estimated value of the relative transverse speed 
between RXTE and the shadow obtained from diffraction pattern fitting 
and the knowledge of RXTE's velocity at the event epoch 
(given in Table~\ref{epoch}), 
one may infer the distance to the occulting body 
by assuming its orbital parameters. 
This assumption is needed because of the
size-distance degeneracy, as one can see that 
$u_{\rm t}$ can be determined from diffraction pattern fitting 
only in units of Fresnel scale per second and the Fresnel scale 
is distance dependent.
Since most of MBAs and TNOs have low-eccentricity orbits,
for simplicity and as an approximation, 
we consider only circular orbits in this study, 
but allow different orbital inclination. 
In Figure~\ref{dist}, 
the relative speed transverse to the Sco X-1 direction 
between RXTE and the occulting body as a function of heliocentric distance 
of the occulting body is plotted, 
and so is the relative speed determined 
from diffraction pattern fitting, 
which is translated from the form of 
Fresnel scale per second into real speed as a function 
of heliocentric distance of the occulting body. 
The intersection of these two curves gives 
the estimate of the heliocentric distance of the occulting body. 
From that distance, one may infer the diameter of the occulting body 
from the best estimated value of the diameter to Fresnel scale ratio $\alpha$, 
also obtained from the diffraction pattern fitting. 

From Figure~\ref{dist}, considering the curve for the minimum inclination
(the thick solid line in Figure~\ref{dist}),
we obtain that the heliocentric distance and diameter of the occulting body
are $r=3.4^{+24}_{-2.4}$ AU and $D=38^{+26}_{-37}$ m.
The uncertainty quoted here is based on the uncertainty of the diffraction
pattern fitting. 
Because the fitting does not provide constraint for high $u_{\rm t}$,
we do not have corresponding constraint 
at the small distance end in Figure~\ref{dist}.
In fact, a very nearby object of meter size 
and with speed at a few kilometers per second
could produce a light curve like Event E1. 
This corresponds to high $u_{\rm t}$ and high $\alpha$
because of the very small Fresnel scale at a very nearby distance.
We therefore artificially assign the lower uncertainty level to
$r$ and $D$ in such a way to symbolically indicate such a situation.
On the other hand, the curve for the largest relative speed 
(the thick dashed line) in Figure~\ref{dist} at about 40 AU
corresponds to an inclination angle of 172$^\circ$ (retrograde orbits).
If we consider high-inclination orbits, 
within the 3-$\sigma$ fitting uncertainty, 
the occulting body could be at about 40 AU (and therefor about 150 m in size).
Of course, such orbits are rare.

Because of the size-distance degeneracy and the weak constraints that 
diffraction pattern fitting to the Event-E1 light curve can provide,
we are not able to draw firm conclusions on the distance and size of the
occulting body causing Event E1.
The size-distance degeneracy may be resolved 
by the knowledge of the occulting-body orbit, which is usually not available
and needs to be assumed,
or by an array of detectors, which, although not yet available now,
 may provide information of the shadow
size directly.
In the following sections we discuss what constraints we may have
on the MBA and TNO size 
distributions based on the search over the 240-ks RXTE/PCA data of Sco X-1.
The result is plotted in Figure~\ref{sizedis}.
We note that, limited by the RXTE/PCA count rate of Sco X-1, which is about
$2\times 10^4$ counts per second per PCU, 
our dip-event search algorithm can detect occultation events
caused by objects of size only down to one or two times Fresnel scale
\citep{chang07}.  
\begin{table*}
\begin{center}
\begin{tabular}{lll}
\hline
 Epoch & RXTE location (km) & RXTE velocity (km/s) \\ 
\hline
 MJD 54622.397997210 
 & $X=-3.947533\times10^7$ & $V_X=+3.482881\times10^1$ \\
 (2008-06-05) 
 & $Y=-1.338443\times10^8$ & $V_Y=-1.043335\times10^1$ \\
  
 & $Z=-5.803026\times10^7$ & $V_Z=-9.082142\times10^{-1}$ \\
\hline
\end{tabular}
\end{center}
\caption{Spacecraft information for Event E1. 
The RXTE location and velocity at the
event epoch 
are expressed in the celestial equatorial coordinate system.
The X-axis is in the direction of the vernal equinox,  
the Z-axis is in the north celestial 
pole direction, and the solar system barycenter is at the origin.
These values are obtained from RXTE 
housekeeping data and the ephemeris of the Earth 
(JPL Horizons On-Line Ephemeris System,
http://ssd.jpl.nasa.gov/horizons.html).
The direction of Sco X-1 is R.A. 
$=16^h 19^m 55.07^s$, Dec.$=-15^\circ 38' 25.0''$ 
and its solar elongation at the event epoch is $169.3^\circ$.
} 
\label{epoch}
\end{table*}  

\section{The MBA size distribution}

Direct observation of the main belt asteroids has only been achieved
down to the size of about 400 meters. 
Earlier surveys revealed a wavy size distribution of MBAs with a bump around
100 km and a second one around 5 km
(see Fig.\ 4 in \citet{cheng04} and Fig.\ 14 in  \citet{obrien05}
for a compiled distribution; 
see also \citet{tedesco02}
for the ISO result, 
\citet{ivezic01} for SDSS, 
and \citet{yoshida07} for Subaru survey). 
It appears that most of the results are consistent with 
a power index of -3 (5 - 40 km) and
of -1.3 (0.4 - 5 km) for the cumulative distribution.
An estimate of the cumulative number of MBAs, based on the ISO result,
is $(1.2\pm 0.5)\times 10^6$ for MBAs with diameter larger than 1 km.

To estimate the occultation event rate, we formulate the MBA size distribution 
as the following:
\beq
N_{>D} = N_i \left(\frac{D}{D_i}\right)^{-b_i}
\,\, ,
\eeq
with $D_1= 5$ km, 
$b_1=3$, $N_1=1.5\times 10^5$ for  5 km $< D <$ 40 km,
$D_2=1$ km, 
$b_2=1.3$, $N_2=1.2\times 10^6$ for  0.4 km $< D <$ 5 km,
and $D_3=0.4$ km,
$b_3=2.5$, $N_3=3.9\times 10^6$ for  $ D <$ 0.4 km.
In this formulation, the normalization is set according to the ISO result, and
the power index 2.5 for the state of collisional equilibrium  with
a constant tensile strength is assumed
for $D<$ 0.4 km.
The expected event rate can be estimated as
\beq
R_{\rm e}=\frac{1}{\tau}\frac{\Omega_\tau}{\Omega_{\rm A}}
\,\,\, ,
\eeq
where
$\Omega_\tau$ is the sky area swept by all MBAs in the time interval of $\tau$,
and $\Omega_{\rm A}$ is the sky area over which MBAs are distributed.
We then have, for MBAs with diameter larger than 10 m,
\beqary
\Omega_\tau & = & \int_{10{\rm m}}^{\infty}
\frac{1}{\triangle^2}\frac{\rd N}{\rd D}\,Dv\tau\,\rd D \\
 & = & \langle\frac{v}{\triangle^2}\rangle\tau
\times 6.6\times 10^8 {\rm km} \,\, ,
\eeqary
where typical values of $v$, the relative speed, and $\triangle$,
the distance from the earth to the MBA, need to be assumed,
and about 99.3\% contribution comes 
from those with diameter smaller than 0.4 km.
Adopting $\triangle=2$ AU and $v=$ 12.6 km/s, we have
\beq
\Omega_\tau=3.7\times 10^{-8} \,\tau \,\, {\rm s}^{-1}
\,\, .
\eeq

The ecliptic latitude distribution of about 13,000 MBAs with $r*<21.5$ 
is shown in Fig.\ 12 of \citet{ivezic01}.
Considering that Sco X-1 is at $5.5^\circ$ north of the ecliptic,
we adopt an equivalent sky coverage area as being spanning over 
$20^\circ$ around the ecliptic, that is,
\beq
\Omega_{\rm A}\approx
360\times 20\times\left(\frac{\pi}{180}\right)^2
\approx 2.2
\,\, .
\eeq
The expected event rate, $R_{\rm e}$
is therefore about $1.7\times 10^{-8}$ s$^{-1}$.

If we consider that Event E1 is caused by an MBA, 
the measured event rate with one detection (Event E1) is 
$R_{\rm m}=(1/240$ ks)$\approx 4.2\times 10^{-6}$ s$^{-1}$.
$R_{\rm m}$ is about 250 times larger than $R_{\rm e}$.
However, if that is the case, the MBA causing Event E1 is about 40 m in size.
The chance is low that one 40-m object is found but no smaller ones,
unless the number of MBAs drops significantly for size smaller than 40 m.
If so and still assuming a single power law for size smaller than 400 m,
 to have one detection in 240 ks for MBAs down to 40 m in size, 
the power index $b_3$ will need to be 5.8. 
It is a very steep power law, as shown in Figure~\ref{sizedis},
and also requires the paucity of smaller MBAs.
We consider this possibility not plausible.

If Event E1 is not caused by an MBA, 
we have no detection of MBAs in the 240-ks data.
Our dip-event search algorithm can detect occultation events
caused by objects of size only down to one or two times Fresnel scale
\citep{chang07}.  
For the MBA distance, it is about 10 m. 
Based on the non-detection, we estimate the upper limit to the size
distribution at the level of assuming one detection.
As calculated above, this level will give a detection rate
250 times larger than $R_{\rm e}$.
To have such a detection rate, if we still adopt a single power law
for the size range between 10 m and 400 m, the power index
$b_3$ needs to be 4.0, instead of 2.5.
This is quite steep, 
although a power index larger than 2.5 is usually expected.
In the collisional equibrium, one expects to have
\citep{obrien03}
\beq
b_3+1=\frac{7+s/3}{2+s/3}
\,\, ,
\eeq
where $s$ is the power index of the strength-size relation.
For $b_3=4.0$, we have $s=-2.25$, which is larger than all the results reported
in the literature to date.

\section{The TNO size distribution}

We describe the TNO size distribution in terms of 
the differential surface density at ecliptic. 
More specifically, we plot in this paper the quantity
$\frac{\rd n}{\rd \log s}$ versus $s$, where
$\rd n$ is the differential number of TNOs {\bf per square degree}
 of size between $s$ and $s+\rd s$ 
and $s$ is the diameter.

\subsection{RXTE observations}

For a background point source, the event rate is
\begin{equation}
\frac{N}{T}=\frac{\int_{s_1}^{s_2}\left(\frac{{\rm d}n}{{\rm d}s}\right)
sv\,{\rm d}s}{d^2}\times(\frac{180}{\pi})^2 \,\,\,\, ,
\end{equation}
where $N$ is the number of detected events 
(assuming a 100\% detection efficiency), $T$ the total exposure time,
$\left(\frac{{\rm d}n}{{\rm d}s}\right)$ the differential size distribution
(in terms of surface density),
$v$ the typical sky-projection relative speed, 
 and $d$ the typical distance to the TNOs.

To derive  
$\left(\frac{{\rm d}n}{{\rm d}s}\right)$ from the event rate, a functional
form of the distribution needs to be assumed. 
On the other hand, if the integration is only over a small range of the size,
we may derive an average value at that size.
Noting that 
$\frac{{\rm d} n}{{\rm d}\log s}=\frac{{\rm d} n}{{\rm d}s}\,s\,\ln 10$,
we have
\begin{equation}
\int_{s_1}^{s_2}\left(\frac{{\rm d}n}{{\rm d}s}\right)
sv\,{\rm d}s
=
\left(\frac{{\rm d}n}{{\rm d}\log s}\right)_{s_1<s<s_2}\frac{v(s_2-s_1)}{\ln 10}
\,\,\, ,
\end{equation}
and
\begin{equation}
\left(\frac{{\rm d}n}{{\rm d}\log s}\right)_{s_1<s<s_2}
=
\frac{d^2}{v(s_2-s_1)}\frac{N}{T}\,\ln 10\,(\frac{\pi}{180})^2
\,\,\, .
\end{equation}

Assuming a typical distance $d=43$ AU and 
a typical relative sky-projection speed
$v=30$ km/s, 
with $T=240$ ks and setting one detection as the upper limit, 
we have in the size range from 60 m to 300 m 
\begin{equation}
\left(\frac{{\rm d}n}{{\rm d}\log s}\right)
< 1.7\times 10^{10}\,\,\mbox{deg}^{-2}
\,\,\, .
\end{equation}

The ecliptic latitude distribution of TNOs is not yet well determined.
To be more precise, one should discuss the latitude 
with respect to the Kuiper Belt Plane (KBP) 
\citep{elliot05}
rather than to the ecliptic,
but the difference is not large.  
Sco X-1 is 5.5$^\circ$ north of
the ecliptic and 4.9$^\circ$ noth of the KBP. 
The CFHT survey 
\citep{trujillo01} reported a 20$^\circ$ half-angle 
of an assumed gaussian distribution in orbital inclination
(with respect to the ecliptic). 
This half-angle in the inclination distribution
translates to about 12.8$^\circ$ for a corresponding half-angle 
in the apparent ecliptic latitude distribution, assuming
circular orbits. 
With this distribution, a factor of 1.14 should be applied to 
convert the estimate at the Sco X-1 latitude to the ecliptic.
However, if we adopt the distribution reported in \citet{elliot05},
a factor of 4 should be applied for conversion to the KBP.
In this paper we use the latitude distribution of \citet{trujillo01}.   


\subsection{TNOs larger than 100 km}

For large TNOs, we use a recent result reported in 
\citet{fuentes09}, in which a double-power-law model describes well
the luminosity function down to $R$-band magnitude of about 27.
With the notation employed in that paper, 
the differential luminosity function reads
\beq
\sigma(R)=C(10^{-\alpha_1(R-23)}
+10^{(\alpha_2-\alpha_1)(R_{\rm eq}-23)}10^{-\alpha_2(R-23)})^{-1}
\eeq
and
\beq
C=\sigma_{23}(1+10^{(\alpha_2-\alpha_1)(R_{\rm eq}-23)})
\,\, ,
\eeq
where $\sigma(R)$ is, with our notation here, equal to $-\frac{\rd n}{\rd R}$.
The reported best fit values of the parameters in that distribution are
$\alpha_1=0.73^{+0.08}_{-0.09}$,
$\alpha_2=0.20^{+0.12}_{-0.14}$,
$R_{\rm eq}=25.0^{+0.8}_{-0.6}$,
and
$\sigma_{23}=1.46^{+0.14}_{-0.12}$.
$C$ is therefore equal to 1.59.
To convert the luminosity function into the size distribution, we assume
a constant albedo and the same distance for all TNOs, and note that
\beq
s=10^{-(\frac{R-R_{\rm eq}}{5})}s_{\rm eq}
\,\, .
\eeq
It follows that
\beqary
\frac{\rd n}{\rd s} & = & \sigma(R)
    \frac{5}{\ln 10}\frac{1}{s_{\rm eq}}10^\frac{R-R_{\rm eq}}{5} \nonumber \\ 
  & = & \frac{5}{\ln 10}\frac{C}{s}10^{\alpha_1(R_{\rm eq}-23)}
        \left[\left(\frac{s}{s_{\rm eq}}\right)^{5\alpha_1}
        +\left(\frac{s}{s_{\rm eq}}\right)^{5\alpha_2}\right]^{-1} 
\,\, , 
\eeqary
and  
\beqary
\frac{\rd n}{\rd \log s} & = & \frac{\rd n}{\rd s} s \ln 10 \nonumber \\ 
  & = & 5C10^{\alpha_1(R_{\rm eq}-23)}
        \left[\left(\frac{s}{s_{\rm eq}}\right)^{5\alpha_1}
        +\left(\frac{s}{s_{\rm eq}}\right)^{5\alpha_2}\right]^{-1}
\,\, . 
\eeqary
Adopting $s_{\rm eq}=90$ km at $R_{\rm eq}=25$, 
which assumes an albedo of 4\% and a distance of 42 AU
and adopts $m_R=-27.6$ for the $R$-band magnitude of the Sun
(\citet{fuentes09}, see also
\citet{fraser09}), 
we have, at $s=90$ km, that
\beq
\frac{\rd n}{\rd \log s}=1.1\times 10^2\,\, \mbox{deg}^{-2}
\,\, .
\eeq
 

\subsection{TNOs of size at 500 m (observations of HST/FGS)}

\citet{schlichting09} reported the detection of 
an occultation event by a TNO of 500 m diameter in the HST/FGS archival data.
They suggest a power index $q=3.9\pm 0.03$ for 
the TNO differential size distribution below 90 km and
the accumulated number of TNOs larger than 500 m is
$2.1^{+4.8}_{-1.7}\times 10^7$ deg$^{-2}$.
These values can be converted into $(\frac{\rd n}{\rd \log s})$ 
in the following way.  
The accumulated distribution is
\beq
\Sigma_{>s}=C_1\left(\frac{s}{s_0}\right)^{1-q}
\,\, ,
\eeq
and
\beqary
\frac{\rd n}{\rd \log s} & = & \frac{\rd n}{\rd s}\,\, s\ln 10 \nonumber \\
 & = & (1-q)C_1\left(\frac{s}{s_0}\right)^{-q}\frac{s}{s_0}\,\ln 10 
\,\, .
\eeqary
For $s_0=500$ m, $C_1=
2.1^{+4.8}_{-1.7}\times 10^7$ deg$^{-2}$.
We therefore have at $s=500$ m that
\beq
\frac{\rd n}{\rd \log s} 
=1.4^{+3.2}_{-1.1}\times 10^8\,\, \mbox{deg}^{-2}
\,\, .
\eeq

\section{Summary and Discussion}

\begin{figure}
\epsfxsize=8.4cm
\epsffile{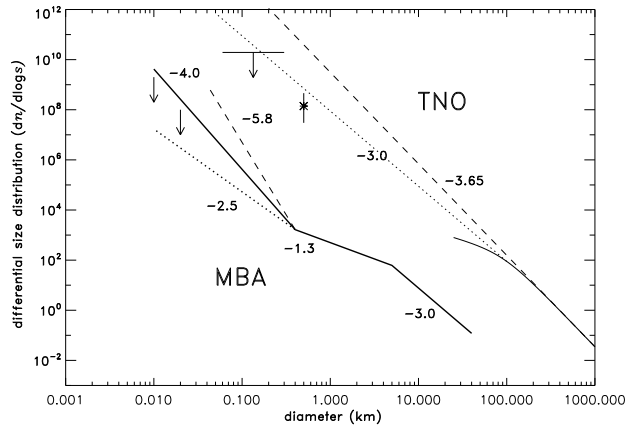}
\caption{Size distributions of main-belt asteroids (MBAs) 
and trans-Neptunian objects (TNOs). 
Plotted here is the differential sky surface density 
at the ecliptic per decade of size in units of number per square degree. 
For TNOs, the upper limit at about 0.1 km, 
denoted with a downward arrow, is derived from our non-detection 
in the 240-ks RXTE/PCA data of Sco X-1, 
and is set at the level of one detection in 240 ks. 
The asterisk symbol at 0.5 km is based 
on the reported HST/FGS detection of an occultation event\citep{schlichting09}.
The thin solid curve is the double-power-law distribution of large TNOs
\citep{fuentes09}, which shows a break at 90 km. 
The HST/FGS result supports the existence of such a break, 
so does our current result. 
The thin dashed line is a direct extrapolation 
from the large size end of the double-power-law towards 
smaller size and the thin dotted line is a power law 
anchoring on the double-power-law at 90 km with a power index of -3.0. 
We note that all the power indices in this figure, 
explicitly printed close to the corresponding power-law lines, 
are for the differential size distribution per decade of size, 
and are therefore the same as that of the cumulative size distribution. 
Upper limits derived from other observations are all above the thin dashed
line and can be found in \citet{schlichting09}.
The double-power-law and the HST/FGS result in fact 
indicate a wavy shape for the TNO size distribution, 
which is similar to the behaviour of MBAs, 
plotted in the lower left half in this figure. 
The thick lines with power indices $-1.3$ and $-3.0$ are for MBAs 
larger than 0.4 km, which are directly observable.
The MBA size distribution has a wavy shape with two bumps at about 100 km 
(not shown in this figure) and 5 km. 
The thick dotted line with a power index of $-2.5$ is for 
a size distribution in collisional equilibrium 
assuming a constant structure strength of MBAs in that size range. 
The thick dashed line with a power index of $-5.8$ is the one implied 
by Event E1 being caused by a 40-m MBA.
The thick solid line with a power index of $-4.0$ is the upper limit 
at the level of assuming one detection down to 10 meters.
}
\label{sizedis}
\end{figure}

In the 240-ks RXTE/PCA data of Sco X-1 taken from June 2007 to October 2009,
we found that millisecond dip events in the light curve are related to
VLEs, which in turn suggests an instrumental origin for those millisecond dips.
The relation is strongest for Type-A VLEs and weakest for Type D. 
One dip event (denoted as Event E1 in this paper)
at random probability of less than $7.5 \times 10^{-5}$ 
is found to be unrelated to any VLEs.
The slight asymmetry, which is similar to many of the VLE-related dips,
and the lacking of side lobes in the light curve of Event E1
cast some doubts on its nature being an occultation event.
However, since the count number is small (the central 4 bins in the light curve
plotted in Figure~\ref{lc} have 2, 0, 5, and 2 counts respectively),
one probably should treat the light curve with high uncertainty.
It is also because of the low count number, 
diffraction pattern fitting does not provide strong enough information
to constrain the property of the occulting body if
it is indeed an occultation event, although acceptable fittings can be
easily obtained.
As discussed in previous sections, 
it might be due to a TNO of 150-m size, but with a rare retrograde orbit,
or, it might be due to an MBA of 40-m size, but the associated detection rate
is incredibly high, or, it might be due to a very nearby object 
of meter size moving at a relative speed of a few kilometers per second.
Given all these uncertainties, we proceed to estimate the upper limits
to the size distribution of TNOs and MBAs 
at the level of assuming one detection.  

The TNO size distribution has been studied in several previous works,
e.g., \citet{bianco10,schlichting09,fraser09,fuentes09,bickerton08,liu08,
roques06,bernstein04}.
In this paper we provide the most updated upper limit at the hectometer size,
which is a size range that optical occultation surveys are unable to explore.
Readers are referred to the caption of 
Figure~\ref{sizedis} for more discussion.
 
The MBA size distribution has been studied by several surveys
\citep{ivezic01,tedesco02,yoshida07} 
down to the size of 400 meters. 
For smaller ones, if the collisional equilibrium 
and a constant tensile strength are assumed, 
a power law with power index -2.5 (cumulative) is expected
\citep{dohnanyi69,obrien03}. 
However, the tensile strength is known to be size dependent, 
and the size distribution for the size range in 
the so-called strength-scaled regime in collisional equilibrium 
is expected to be steeper than the constant-strength solution
\citep{benz99,obrien03}. 
The upper limit that we estimate,
as shown in Figure~\ref{sizedis}, 
has a  power index $-4.0$, 
which in turn suggests a power index of $-2.25$ 
for the strength-size relation
\citep{obrien03}, 
expressed in the form of the critical specific energy 
(conventionally denoted with $Q^*_D$) as a function of size. 
This is still much steeper than all the reported strength-size relation 
in the literature to our knowledge, 
of which the steepest one has a power index of about $-1$
\citep{durda98}. 

Studying population properties of small TNOs and MBAs with serendipitous
occultation events in X-rays is still in its infant stage,
currently only possible for RXTE/PCA, the X-ray instrument with the 
largest effective area, observing Sco X-1, 
the brightest X-ray source in the sky. 
It, however, suffers significantly from the instrumental dead time caused by
high energy events. 
Observations with 
RXTE and future facilities, 
such as ASTROSAT/LAXPC, AXTAR\citep{chakrabarty08}, and IXO, 
together with a more rigorous design of the survey, similar to 
those considered in the optical band
\citep{bickerton09,nihei07,gaudi04},
will be able to yield more reliable and complete
information, particularly 
when more X-ray sources in different directions can be 
employed. 

\section*{Acknowledgments}
We thank Ed Morgan, who designed the new RXTE/PCA data mode 
and coordinated the new RXTE observations of Sco X-1, 
and Jau-Shian Liang, who helped to identify RXTE velocity 
in the solar system at the event epochs. 
We appreciate very much Steve Bickerton's comments, which 
significantly improved this paper.
We are also grateful to Francoise Roques, Alain Doressoundiram, 
and Sun-Kung King for helpful discussion. 
This work was supported by the National Science Council of 
the Republic of China under grant NSC 96-2628-M-007-012-MY3.

\label{lastpage}
\end{document}